\pdfoutput=1
\documentclass[10pt, a4paper, final]{article}

\usepackage{lrec}
\newcommand*{\citep}[1]{\cite{#1}}
\newcommand*{\citet}[1]{\newcite{#1}}

\usepackage{titlesec}
\renewcommand{\thesection}{\arabic{section}}
\renewcommand{\thesubsection}{\thesection.\arabic{subsection}}

\titlelabel{\thetitle.\quad}

\usepackage[utf8]{inputenc}
\usepackage{mathptmx}

\usepackage{csquotes}
\usepackage{xstring}
\usepackage[binary-units]{siunitx}
\usepackage{graphicx}
\usepackage{subcaption}

\usepackage{tabularx}
\usepackage{booktabs}
\usepackage{makecell}

\usepackage[hyphens]{url}
\usepackage[hidelinks, breaklinks, pdfusetitle]{hyperref}
\usepackage[all]{hypcap}
\usepackage[capitalise, nameinlink, noabbrev]{cleveref}

\usepackage[obeyFinal, colorinlistoftodos]{todonotes}
\usepackage{acro}
\DeclareAcronym{HMM}{long=hidden Markov model}
\DeclareAcronym{GUI}{long=graphical user interface}
\DeclareAcronym{MFCC}{long=mel-frequency cepstral coefficient}
\DeclareAcronym{PLP}{long=Perceptual Linear Prediction}
\DeclareAcronym{LD}{long=Levenshtein distance}
\DeclareAcronym{MFA}{long=Montreal Forced Aligner}
\DeclareAcronym{PER}{long=phone error rate}
\DeclareAcronym{WER}{long=word error rate}
\DeclareAcronym{TSCC}{long=TechSmith Screen Capture Codec}
\DeclareAcronym{HID}{long=human interface device}

\title{A Multimodal Corpus of Expert Gaze and Behavior\\ during Phonetic Segmentation Tasks}

\name{%
  Arif Khan\textsuperscript{1--3},
  Ingmar Steiner\textsuperscript{1,2},
  Yusuke Sugano\textsuperscript{4},
  Andreas Bulling\textsuperscript{1,5},
  Ross Macdonald\textsuperscript{6}
}
\hypersetup{pdfauthor={Arif Khan, Ingmar Steiner, Yusuke Sugano, Andreas Bulling, Ross Macdonald}}

\address{%
  \textsuperscript{1}Multimodal Computing and Interaction, Saarland University, Germany, \\
  \textsuperscript{2}German Research Center for Artificial Intelligence (DFKI GmbH), Saarbrücken, Germany, \\
  \textsuperscript{3}Saarbrücken Graduate School of Computer Science, Germany, \\
  \textsuperscript{4}Osaka University, Japan \\
  \textsuperscript{5}Max Planck Institute for Informatics, Saarbrücken, Germany \\
  \textsuperscript{6}University of Manchester, UK, \\
  \{arifkhan,steiner\}@coli.uni-saarland.de
}

\abstract{
Phonetic segmentation is the process of splitting speech into distinct phonetic units.
Human experts routinely perform this task manually by analyzing auditory and visual cues using analysis software, which is an extremely time-consuming process.
Methods exist for automatic segmentation, but these are not always accurate enough.
In order to improve automatic segmentation, we need to model it as close to the manual segmentation as possible.
This corpus is an effort to capture the human segmentation behavior by recording experts performing a segmentation task.
We believe that this data will enable us to highlight the important aspects of manual segmentation, which can be used in automatic segmentation to improve its accuracy.
\\ \newline  }

\begin{document}
\setcounter{page}{4277}

\maketitleabstract

\section{Introduction}
\label{sec:introduction}

\begin{figure*}
  \begin{subfigure}{\columnwidth}
    \includegraphics[width=\linewidth]{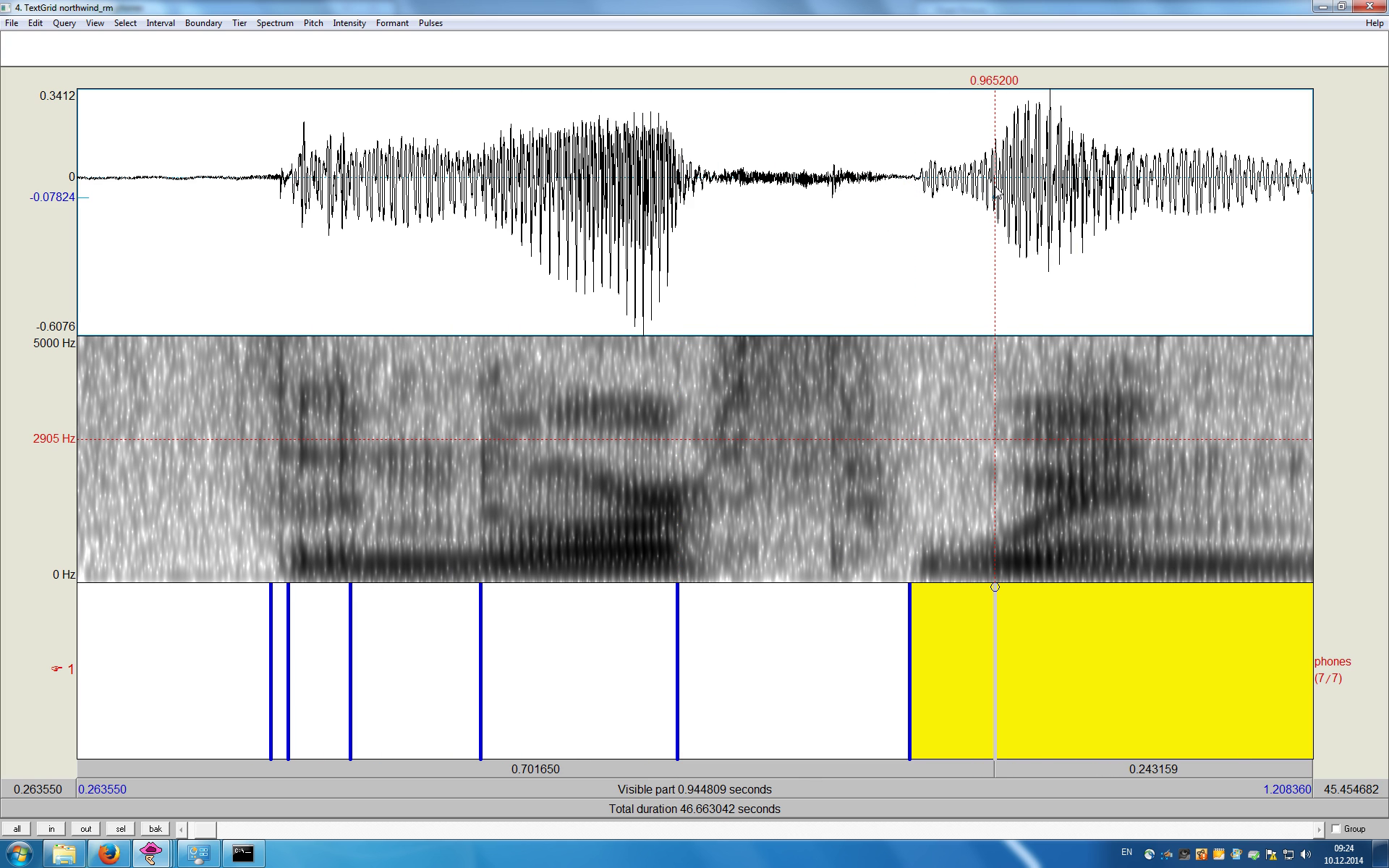}
    \caption{A screenshot of a sound recording and annotation in Praat.
      The \ac{GUI} is split into three sections: waveform (top), spectrogram (middle), and annotation (bottom).}
    \label{fig:screencapture}
  \end{subfigure}
  \hfill
  \begin{subfigure}{\columnwidth}
    \frame{\includegraphics[width=\linewidth]{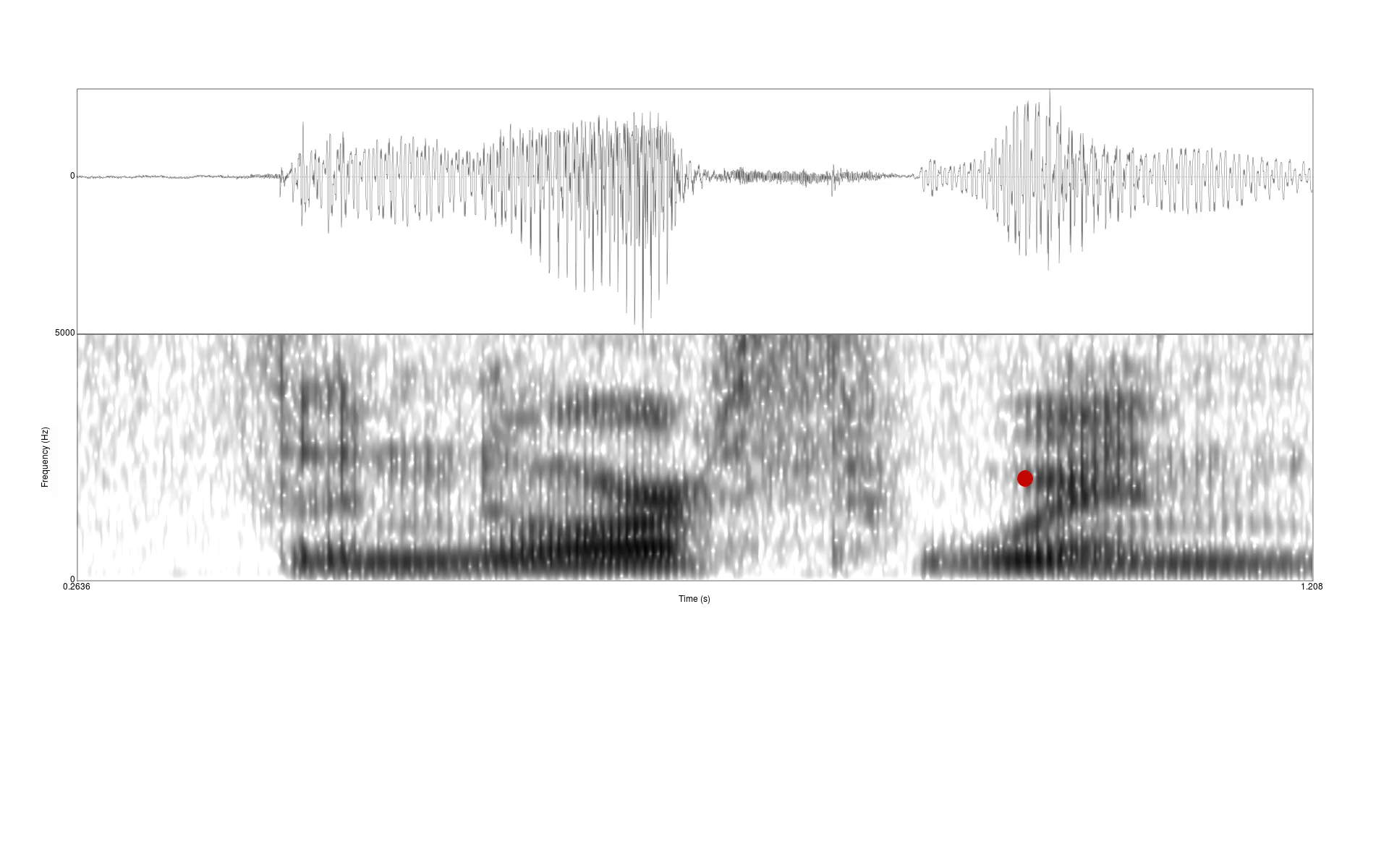}}
    \caption{Corresponding scene data reconstructed from recorded audio using Praat log and gaze information.
      Here, the subject is looking at a formant in the spectrogram; the fixation is rendered in red.}
    \label{fig:gazereconstruction}
  \end{subfigure}
  \caption{One frame from the screen capture video (left) and the corresponding reconstruction (right).}
  \label{fig:screenshots}
\end{figure*}

Speech segmentation is the process of splitting the acoustic speech signal into distinct units by placing timestamped boundaries.
This forms a crucial data processing step for phonetic analysis, as well as speech technology applications such as text-to-speech synthesis and automatic speech recognition.
The results and output quality depend on accurately segmented speech data.

Speech segmentation can be done manually, using specialized software, e.g., \emph{Praat} \citep{Boersma2001}, \emph{Wavesurfer} \cite{Beskow2000}, \emph{ELAN} \cite{Sloetjes2008}, and \emph{EMU} \cite{Winkelmann2017}.
In this workflow, a speech recording is displayed as a waveform and/or spectrogram, and boundaries are inserted using the mouse or keyboard (cf.\ \cref{fig:screencapture}).
Short audio segments can be played back to validate the boundary placement.
This process is repeated until the whole audio file is segmented.
Manual segmentation by experts is considered to produce the best phonetic segmentation one can achieve for any given data \citep{svendsen1987automatic,wesenick1996estimating}.
One reason for this is that they combine experience with multiple sources of information.
However, there are some critical drawbacks of manual segmentation which make it impractical for large speech corpora.
The first is that it is very laborious and time consuming;
on average, manual segmentation can take up to \SI{30}{\second} per phone \citep{Leung1984ICASSP,stolcke2014highly} to segment.
As a result, newly recorded speech data cannot be used quickly if manual segmentation is desired.
Secondly, the exact placement of boundaries is subjective, and there may be disagreement between multiple experts.

The second method of segmentation is doing it automatically, by training a model on the audio data, and then using it to segment speech.
In this method, the accuracy of the segmented speech directly depends on the quality of the trained model, which itself depends on the quality of training data.
Previous studies have used different approaches for automatic segmentation.
For a long time, researchers have used \acp{HMM} for automatic segmentation \citep{rabiner1989tutorial,juang1991hidden,toledano2003automatic,brognaux2016hmm}.
Others have used neural networks for automatic segmentation \citep{karjalainen1998efficient,schwarz2006hierarchical}.
One commonality of these approaches is the use of only audio as input features for training the model.
The audio is processed to extract acoustic features, which are then used for training.
Several techniques are available for extracting acoustic features from speech, but
the most commonly used are \acp{MFCC} \citep{logan2000mel} and \ac{PLP} \citep{hermansky1990perceptual}.
While the use of only audio as acoustic features produces acceptable results for most segmentation requirements, humans use more than audio for segmenting speech.
To improve automatic segmentation, we therefore want to add more modalities to model it as closely as possible to the manual segmentation.
We hope that modeling automatic segmentation in this way will produce better results.

To this end, we first need to analyze the human segmentation behavior and highlight the key information sources that humans experts use to segment speech.
Our data includes gaze information, which shows where the experts look on the screen,
the audio to which they listen during segmentation,
video from a webcam attached to the monitor,
and a screen recording of what they are viewing.
To the best of our knowledge, this is the first corpus that records the human segmentation in such a setup.

The rest of this paper is organized as follows.
\Cref{sec:thecorpus} provides details of how the data was recorded, along with the format and structure.
In \cref{sec:analysis}, we present the results of some preliminary analysis conducted on the data.
Finally, the conclusion and future use of the data is outlined in \cref{sec:conclusion}.

\section{The Corpus}
\label{sec:thecorpus}

In order to study the behavior of human experts during speech segmentation tasks, we designed and recorded the multimodal corpus described in this section.

\subsection{Preparation}

We recorded a native speaker of Scottish English, reading aloud the standard passage, \enquote{The North Wind and the Sun} \citep{IPA1999Handbook}.
The recording was made in a sound-attenuated booth, with a close-talking microphone, sampling at \SI{48}{\kilo\hertz} with \SI{24}{\bit} quantization.
The resulting file has a duration of \SI{46}{\second}.
%

\subsection{Data Collection}
\label{sec:data-collection}
\acreset{GUI}

\begin{table*}
  \begin{tabularx}{\linewidth}{XcScSS}
      \toprule
      Subject & Gender & {Age (years)} & Native Language &
      {Experience (years)} & {Segmentation Time (\si{\minute})} \\
      \midrule
      01 & F & 26 & German & 7  & 44 \\
      02 & M & 47 & German & 20 & 55 \\
      03 & M & 37 & German & 15 & 73 \\
      04 & F & 35 & Polish & 10 & 96 \\
      05 & F & 27 & German & 4  & 71 \\
      06 & F & 22 & German & 1.5 & 80 \\
      07 & F & 22 & German & 4 & 92 \\
      \bottomrule
\end{tabularx}
\caption{Age, gender, native language, and segmentation experience of the subjects who participated in the data collection.}
\label{tab:participants}
\end{table*}

\begin{figure*}
  \centering
  \includegraphics[width=\linewidth]{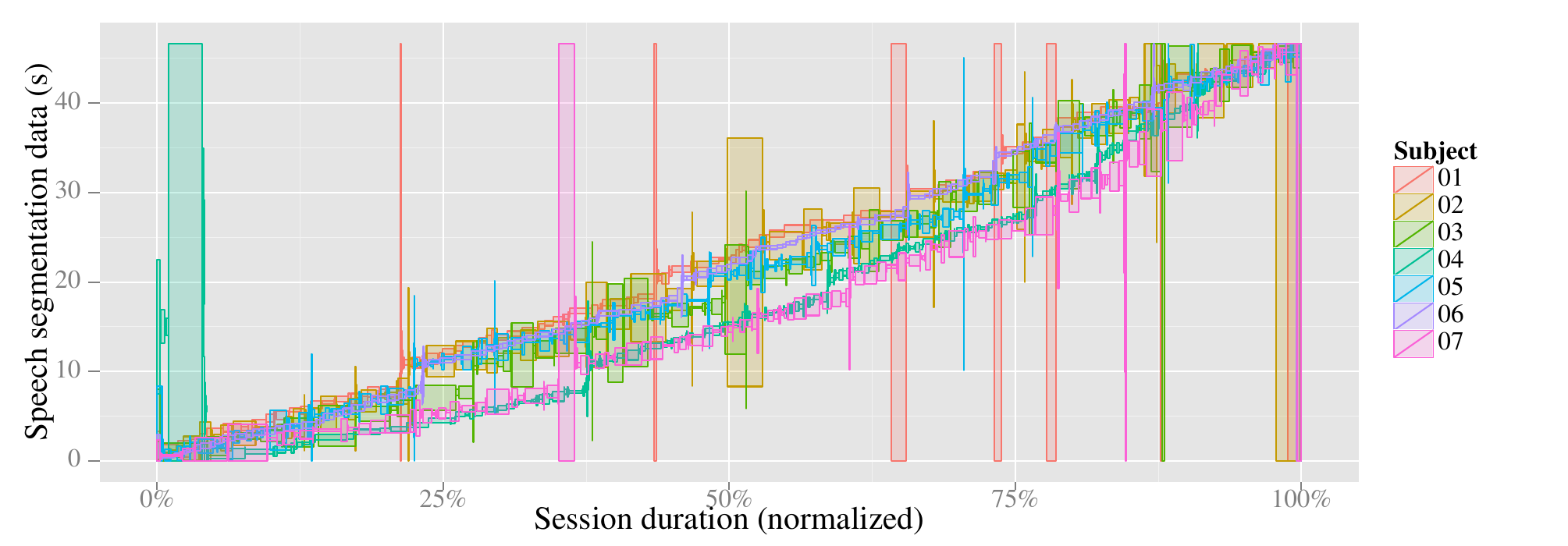}
  \caption{Speech segmentation data spans which were viewed as scenes over the (normalized) duration of the segmentation task.
  Each rectangle represents the portion of time (rectangle width) spent segmenting a span of recorded speech, while the rectangle height represents the duration of that span.}
  \label{fig:task_duration}
\end{figure*}

We recorded seven subjects, with the instruction that they were to segment (but not label) the recording into phones using the Praat \ac{GUI}.
All of the subjects who participated in the data collection are trained phoneticians with varying amounts of experience;
details are given in \cref{tab:participants}.

The participants took different amounts of time (\SIrange{44}{96}{\minute}) to complete the task.
The normalized session duration for all the subjects in shown in \cref{fig:task_duration}.
We did not control the speed in which the participants completed the task, so each took time according to his or her preference, which resulted in different session durations.

We used a Tobii TX300 eyetracker,%
\footnote{\url{https://www.tobiipro.com/product-listing/tobii-pro-tx300}}
to record the gaze movements and capture where the subject looked on the computer screen during the entire session, at a sampling rate of \SI{120}{\hertz}.
For each subject, before the beginning of recording, we first calibrated the eyetracker.
The calibration is done to adjust the height of head and seating position, which is different for each subject.
Using the \emph{TobiiStudio} software (v3.2.3), we also recorded the screen content itself (at a resolution of \SI{1920x1200}{pixels}), as well as any audio the subjects played back from the recording during the segmentation task.
In addition to the gaze information and screen recording, TobiiStudio also allowed us to log any keystrokes and mouse clicks during the recording session, as well as the video from a webcam facing the subject, at a resolution of \SI{640x480}{pixels}.
The screen capture and webcam were intended to validate the subjects' head movements and input device logging.

In addition to these modalities, we polled the application state of the Praat \ac{GUI}, once per second, in order to log the zoom level of the audio recording shown and other application-specific data.
Finally, the segmentation itself, produced by each subject over the course of the session, was saved in Praat's widely supported \emph{TextGrid} annotation file format.

\subsection{Data Processing}
\label{sec:data-processing}

After each recording session, the logs from TobiiStudio and Praat were exported to ASCII text files and compressed.
The screen recordings and webcam videos, as well as the audio playback recordings, were exported from TobiiStudio in ASF containers, in \ac{TSCC}, Microsoft Video 1, and MP3 format, respectively, the latter at \SI{22}{\kilo\hertz} and \SI{16}{\bit} quantization, at a bitrate of \SI{128}{\kilo\bit\per\second}.

In order to manipulate the multimedia streams from each recording session more efficiently, we first converted the video to H.264 format (which allowed more robust seeking and reduced the file sizes -- from \SI{52}{\giga\byte} to \SI{3}{\giga\byte} without noticeable loss in quality), transcoded the audio to FLAC format,%
\footnote{\url{https://xiph.org/flac/}}
and multiplexed all three streams into a single Matroska video container,%
\footnote{\url{https://matroska.org/}}
using FFmpeg.%
\footnote{\url{https://ffmpeg.org/}}

Next, we parsed the Praat logs to identify time segments in each recording session during which the subject was viewing the same zoom level and interval of the audio recording;
doing this allowed us to treat them as quasi-static \emph{scenes} viewed by the subject.
The session times as well as the audio recording times of each scene were collected into a YAML file.%
\footnote{\url{http://yaml.org/}}


After determining the constant time offset between the Praat and TobiiStudio logs, we could then select the gaze data related to each scene and store it in a structured format, validating it via the screen recording.
The data is structured by scene and also includes the duration and location (absolute and classified by \ac{GUI} region) of each fixation.
Based on this information, we reconstructed the relevant information in each scene and synthesized it into a second video stream with the gaze location rendered as a red circle (cf.\ \cref{fig:gazereconstruction}).
We also extracted the signal time codes of each scene and added them as a subtitle track.
The resulting YAML files and multimedia streams were finally packaged and provided as a data dependency for analysis.

\section{Analysis}
\label{sec:analysis}

Our initial analysis concerns the eyetracking data.
The main purpose of the corpus was to allow us to analyze the manual segmentation behavior and to identify modalities and features useful for modeling segmentation.
For the analysis of the eyetracking data, it is important to understand the concepts of \emph{fixation} and \emph{saccades}.
If the $\langle x,y\rangle$ location of the gaze on the screen does not change significantly within some time frame, then those gaze events are classified as a fixation.
The movement of gaze between two fixations is referred to as a saccade.
The actual time duration for which the $\langle x,y\rangle$ location movement should remain constant is subjective and device dependent.
We used the default settings of Tobii to identify the fixations and saccades, the details of which are described by \citet{ollson2007real}.

For analysis, the Praat \ac{GUI} on the screen is divided horizontally into three sections, each representing a different portion of the screen.
We refer to these sections as \emph{waveform}, \emph{spectrogram}, and \emph{annotation}, as shown in \cref{fig:screenshots}.
The waveform represents the oscillogram of the audio recording in Praat.
The spectrogram represents the time-frequency-energy representation of the signal;
the $x$ axis represents time, and the $y$ axis, the frequency of the signal, while the grayscale value indicates the energy in each time-frequency bin.
The annotation section is used by the subjects to place the boundaries.
This is the only section which can be edited by the user for creating and manipulating time-aligned annotations (boundaries and labels).

\subsection{Scenes}

Further to the progress visualization in \cref{fig:task_duration}, \cref{tab:scene_details} summarizes the number of scenes the subjects viewed over the course of their session.
As can be seen, subjects 01 to 03 and 05 to 07 used almost the same number of scenes for segmentation.
Subject 04 viewed a larger number of scenes with the second lowest average scene length, indicating that this participant preferred to \enquote{zoom in} more than the others.

\subsection{Fixations}

\begin{table}
  \begin{tabularx}{\linewidth}{XS[table-format=3.0]SS}
      \toprule
      Subject & {\thead{Total\\scenes}} & {\thead{Total\\duration (\si{\second})}} & {\thead{Average\\duration (\si{\second})}} \\
      \midrule
      01 & 157 & 519.79 & 3.31 \\
      02 & 157 & 593.63 & 3.78 \\
      03 & 150 & 562.23 & 3.74 \\
      04 & 522 & 627.67 & 1.20 \\
      05 & 308 & 671.64 & 2.18 \\
      06 & 352 & 361.50 & 1.02 \\
      07 & 276 & 652.49 & 2.36 \\
      \bottomrule
  \end{tabularx}
  \caption{The total number of scenes, sum of scenes length and average scene length the subjects used for segmenting the audio recording.}
  \label{tab:scene_details}
\end{table}

\begin{figure}
  \includegraphics[width=\linewidth]{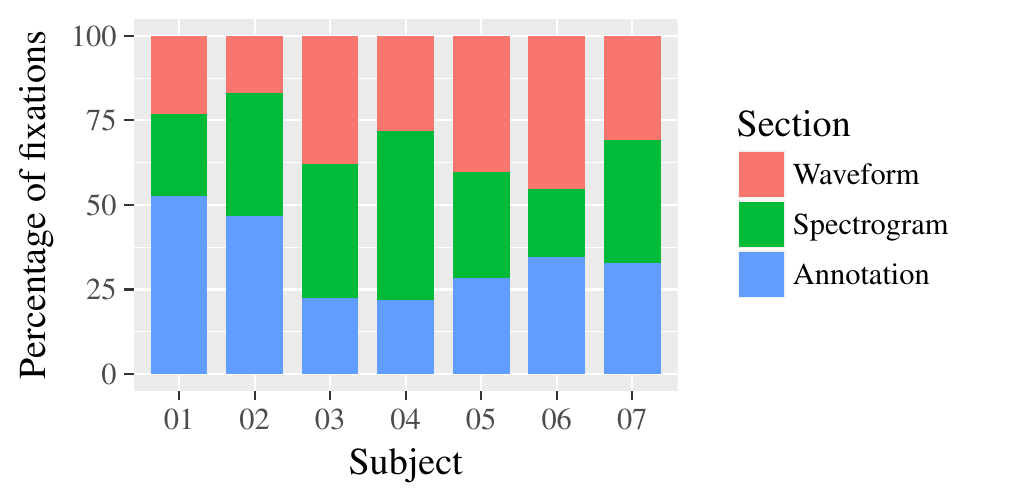}
  \caption{Average fixations for each subject in the three sections of the Praat \ac{GUI}.}
  \label{fig:avg_fixations}
\end{figure}

One of the most important questions is, where the subjects look on the screen during the manual segmentation task.
To answer this question, we calculated the proportion of gaze events in the three sections of the screen.
\Cref{fig:avg_fixations} shows the percentage of fixations in each of the three screen sections for all subjects.
The fixations in the \emph{annotation} area can be disregarded, because in order to place the boundary, the subjects have to carefully \enquote{click} in the right location and during this process, a lot of gaze activity may occur in this section.
The fixations in the \emph{waveform} and \emph{spectrogram} sections are important and have a mixed pattern.
All subjects have a higher number of fixations in the \emph{spectrogram} section than in the \emph{waveform} section.%
\footnote{The exception is subject 06;
  this may be because she had the least amount of segmentation experience (see \cref{tab:participants}) and relied more on the waveform section to segment.}

\section{Conclusion and Outlook}
\label{sec:conclusion}

In this paper, we have presented a multimodal corpus of behavior data from expert phoneticians performing a manual speech segmentation task.
All important information sources that are relevant to the segmentation task were recorded.
This includes gaze, playback audio, video, and screen recording.
The produced segmentation, as well as events logged from the keyboard, mouse, and Praat \ac{GUI} are also provided.
We believe that this data will prove valuable for research in observing and understanding manual segmentation.

This corpus can help identify critical information sources used by humans during manual segmentation, which can be modeled to improve the accuracy of automatic segmentation.
In addition, this data can be useful in analyzing the interaction of phoneticians with speech segmentation software (Praat) and can be used to improve the usability of such a software.
For example it might be possible to modify the way the boundaries are defined or to introduce a software feature to visualize the predicted complexity of speech regions while they are being segmented.

The processed data (cf.\ \cref{sec:data-processing}) has been released under a Creative Commons license (CC-BY-NC-SA) and published on GitHub,%
\footnote{\url{https://git.io/eyeseg-data}}
along with the processing recipes.
This public release \emph{excludes} the webcam videos, in order to protect the privacy of our participants.

\section{Acknowledgements}
We are extremely grateful to all of our participants, who gave their time for the segmentation task and provided valuable feedback.

This study was funded by the German Research Foundation (DFG) under grant number EXC~284.

\section{Bibliography}
\bibliographystyle{lrec}
\bibliography{mybib}

\listoftodos

\end{document}